\documentclass{cs19proc}

\editors{G.~A. Feiden}
\publisher{Zenodo}
\conference{The 19th Cambridge Workshop on Cool Stars, Stellar Systems, and the Sun}
\conferencedate{2016}

\newcommand{\etor}{\langle B_{\rm tor}^2 \rangle}
\newcommand{\epol}{\langle B_{\rm pol}^2 \rangle}
\newcommand{\mdot}{\dot{M}}

\title{Stellar magnetism, winds and their effects on planetary environments}
\author{A.~A.~Vidotto$^{1}$}

\affiliation{$^{1}$School of Physics, Trinity College Dublin, the University of Dublin, Dublin-2, Ireland}

\shorttitle{Stellar magnetism, winds and their effects on planetary environments}
\shortauthors{Vidotto}
\abs{Here, I review some recent works on magnetism of cool, main-sequence stars, their winds and potential impact on surrounding exoplanets. The winds of these stars are very tenuous and persist during their lifetime. Although carrying just a small fraction of the stellar mass, these magnetic winds carry away angular momentum, thus regulating the rotation of the star. Since cool stars are likely to be surrounded by planets, understanding the host star winds and magnetism is a key step towards characterisation of exoplanetary environments. As rotation and activity are intimately related, the spin down of stars leads to a decrease in stellar activity with age. As a consequence, as stars age, a decrease in high-energy (X-ray, extreme ultraviolet) irradiation is observed, which can affect the evaporation of exoplanetary atmospheres and, thus, also altering exoplanetary evolution.} 

\begin{document}
\maketitle

%%%%%%%%%%%%%%%%%%%%%%%%%%%%%%%%%%%%%%%%%%%%%%%%%%%%%%%%
\section{Introduction}
\begin{figure}[ht]
	\centering
	\includegraphics[width=\columnwidth]{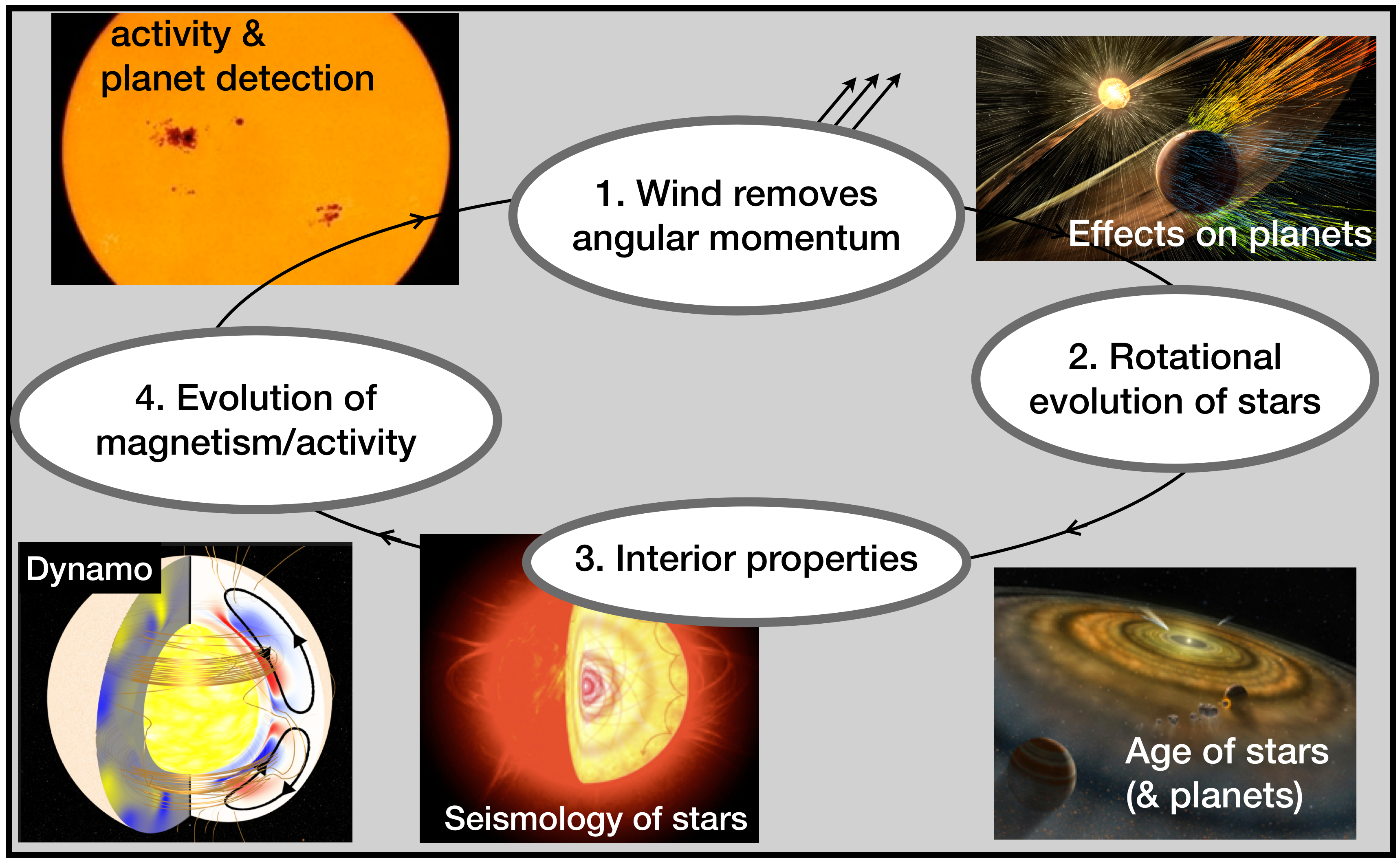}
	\caption{The big picture: evolution of winds of cool dwarf stars. As the star ages, its rotation and magnetism decrease, causing also a decrease in angular momentum removal. In the images, I highlight some of the areas to which wind-rotation-magnetism interplay is relevant. }
	\label{fig:fig_big_picture}
\end{figure}

Cool dwarf stars are believed to lose mass in the form of winds during their entire lifetime. These winds, however, are usually quite tenuous and do not carry away a significant fraction of the stellar mass. Nevertheless, they are fundamental for regulating the rotation of these stars. Because these winds are magnetic in nature, they are able to extract a significant amount of angular momentum from the star. Consequently, the star spins down as it evolves. Because of this variation in surface rotation, there is a redistribution of internal angular momentum transport, which changes the interior properties of the star. With a different internal structure, the dynamo that is operating inside the star changes, changing therefore the properties of the emerging magnetic fields. With a new surface magnetism, the stellar wind also changes and this cycle repeats itself over and over again during the entire lifetime of the star (Figure \ref{fig:fig_big_picture}).

\begin{figure}[ht]
	\centering
	\includegraphics[width=\columnwidth]{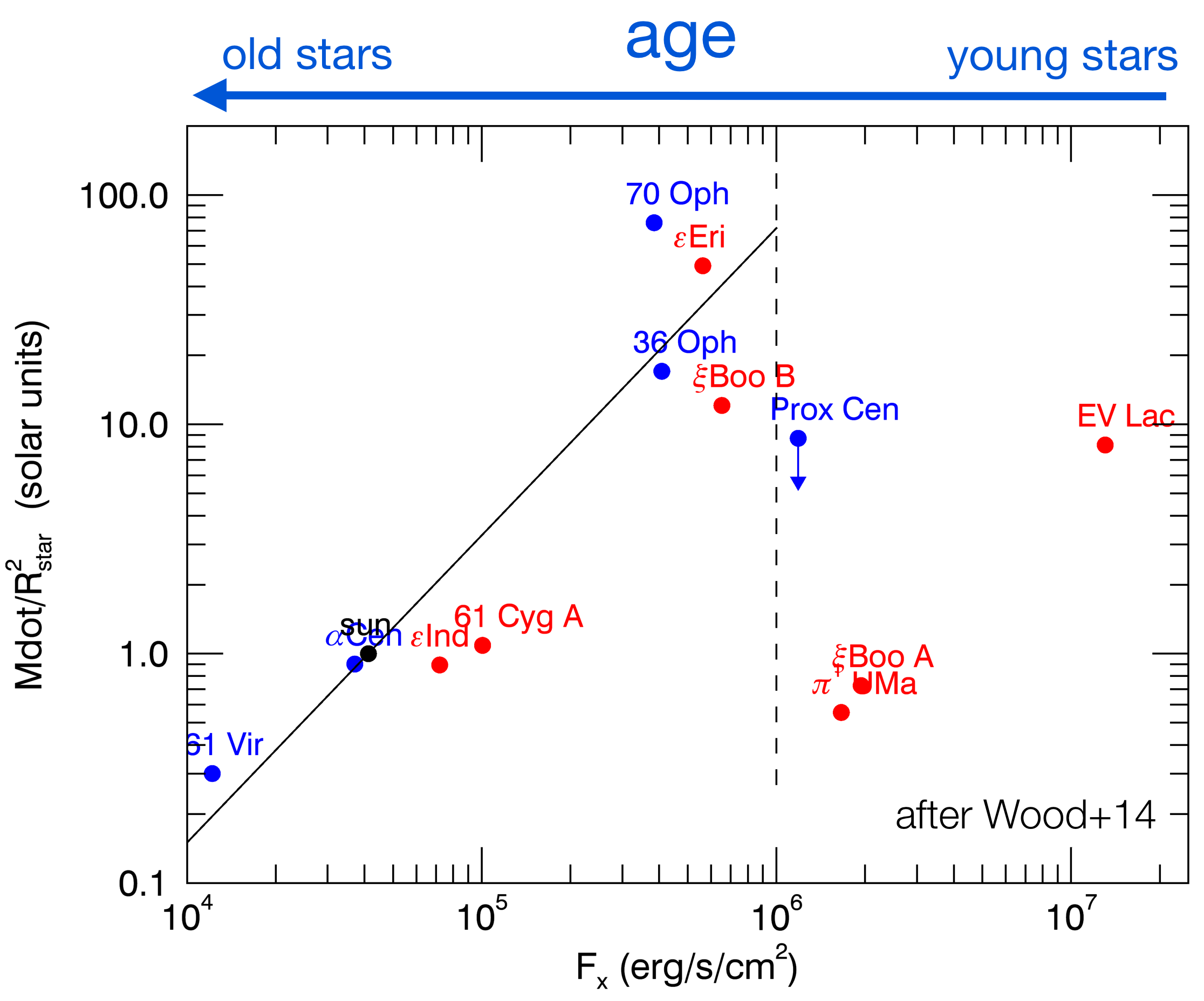}
	\caption{Mass-loss rate per unit area as a function of X-ray flux, based on \citet{2014ApJ...781L..33W}. Because X-ray flux is a measure of stellar activity, it can be used as a rough proxy for age, whereby solar-type stars with large X-ray fluxes are, in general, younger. Red symbols are stars that have reconstructed surface magnetic fields (Section \ref{sec.winds}). }
	\label{fig.wood14}
\end{figure}

Therefore, as the stars ages, its rotation decreases, along with its chromospheric activity  \citep{1972ApJ...171..565S}, magnetism \citep{2014MNRAS.441.2361V} and winds \citep{2004LRSP....1....2W}. Figure \ref{fig.wood14} shows how the mass-loss rate depends on the X-ray flux of the star \citep[adapted from][]{2014ApJ...781L..33W}. Because X-ray flux is a measure of stellar activity, it can be used as a rough proxy for age. Therefore, stars with large X-ray fluxes tend to be younger than stars with more modest X-ray fluxes. What \citet[][and previous works]{2014ApJ...781L..33W} found is that as the star ages, its mass-loss rate tends to decrease. 

\begin{figure*}[!ht]
	\begin{center}
	\includegraphics[width=0.8\textwidth]{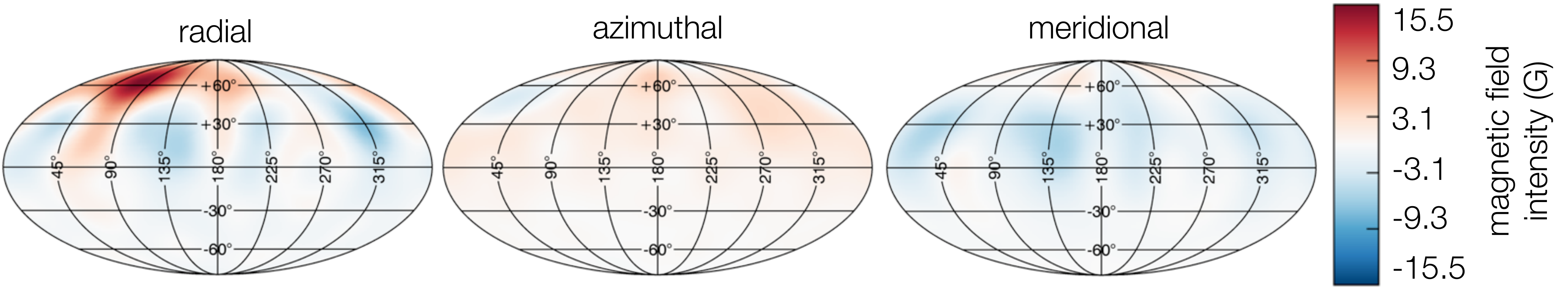}	
	\end{center}
	\caption{The magnetic field of the star $\tau$ Boo at Jun/2007, reconstructed \citep{2008MNRAS.385.1179D} using the ZDI technique. From left to right: the radial, azimuthal (West-East) and meridional (North-South) components.
	\label{fig.magmap}}
\end{figure*}

Some questions I will address in this article are: How has the solar wind evolved? What is the implication of this evolution for young exoplanetary systems? Is the evolution regulated by the topologies of stellar magnetism? 
In Section \ref{sec.magnetism}, I will present the latest results in surveys of imaging of surface magnetic fields. Section \ref{sec.winds} compiles some recent findings of stellar winds and to which extent the magnetic field topology can affect these winds. In Section \ref{sec.planets}, I illustrate some effects that stellar magnetism and winds can have on planetary environments. Finally, in Section \ref{sec.remarks}, I present a summary and some concluding remarks.

%%%%%%%%%%%%%%%%%%%%%%%%%%%%%%%%%%%%%%%%%%%%%%%%%%%%%%%%
\section{Empirical trends on large-scale stellar magnetism}\label{sec.magnetism}
Stellar magnetism can be probed with different techniques. A technique that has been particularly successful in imaging stellar magnetic fields is Zeeman Doppler Imaging (ZDI, \citealt{1997A&A...326.1135D}). Through a series of spectropolarimetric observations aiming at monitoring the star through about a couple of rotation periods, this technique has allowed the reconstruction of the large-scale surface fields of more than one hundred cool stars \citep[e.g.,][]{2006MNRAS.370..629D,2008MNRAS.390..567M,2008MNRAS.388...80P,2009MNRAS.398.1383F,2011MNRAS.413.1922M, 2016MNRAS.459.4325M, 2016MNRAS.457..580F}. Figure \ref{fig.magmap} illustrates an output of this technique, in which the three components of the stellar magnetic fields are reconstructed: radial, azimuthal (West-East) and meridional (North-South). 

With the increase in the number of stars whose magnetic fields have been mapped, several trends have emerged. Here, I highlight a few of them:
\begin{enumerate}
\item \textit{Magnetism decays with age:} By compiling a sample of about 100 ZDI maps, \citet{2014MNRAS.441.2361V} showed that the average magnetic field intensity decays with age$^{-0.65 \pm 0.04}$. This trend is valid over three orders of magnitude in field intensity (from G to kG fields) and four orders of magnitude in ages (from Myr to several Gyr). Although the trend is clear, it also presents a large scatter. Part of this scatter is due to short-term evolution of magnetism, including magnetic cycles \citep[e.g.][]{2008MNRAS.385.1179D,2015A&A...573A..17B,2016arXiv160601032B,2014A&A...569A..79J}.  
\item \textit{For the same age, no abrupt change in magnetic field geometries is found:} Recently, \citet{2016MNRAS.457..580F} reconstructed the surface magnetic field topologies of solar-mass stars in open clusters. One of the  benefits of their approach is that stars belonging to the same cluster are coeval, with well defined ages. In this way, they were able to study how magnetism with stars with similar masses and ages vary as a function of rotation periods. For example, close to the zero-age main sequence, \citet{2016MNRAS.457..580F} found that there is a significant scatter in the magnetic properties, and no abrupt changes in the geometry is found for very fast rotators (with rotation periods $P_{\rm rot} <2$ days) and moderate rotators ($P_{\rm rot} >2$ days)
\item \textit{Toroidal fields appear when the tachocline develops:} Magnetic maps, such as the ones presented in Figure \ref{fig.magmap}, can also be decomposed into poloidal and toroidal fields. \citet{2015MNRAS.453.4301S} showed that stars that possess tachoclines show a distinct magnetic field topology than those stars without a tachocline. The tachocline is essentially an interface separating the radiative core from the convective envelope. Our Sun, which is partially convective, presents a tachocline. As one moves towards lower stellar masses, the convective envelope extends deeper in the star, until at about a mass threshold of about $0.4~M_\odot$ (in the mid-M dwarf regime), the star becomes fully convective and the tachocline no longer exists.  \citet{2015MNRAS.453.4301S} showed the partially convective stars can have a wide range of fraction of toroidal fields, while the fully convective ones were restricted to toroidal fractions smaller than $ \sim 30\%$. This result has also been demonstrated in previous studies concentrating on M dwarfs \citep{2008MNRAS.390..545D,2010MNRAS.407.2269M,2016MNRAS.461.1465H}. Another interesting result from  \citet{2015MNRAS.453.4301S} was that the toroidal and poloidal fields grow together, but at different rates. While in solar-type stars (those possessing tachoclines), the toroidal magnetic energies ($\etor$) relate to the poloidal magnetic energies ($\epol$) as $\etor \propto \etor^{1.25\pm0.06}$, in  fully convective stars this relation is shallower  $\etor \propto \etor^{0.72\pm0.08}$.
\end{enumerate}

%%%%%%%%%%%%%%%%%%%%%%%%%%%%%%%%%%%%%%%%%%%%%%%%%%%%%%%%
\section{Recent studies on magnetised stellar winds}\label{sec.winds}

\begin{figure*}[!ht]
	\begin{center}
	\includegraphics[width=0.8\textwidth]{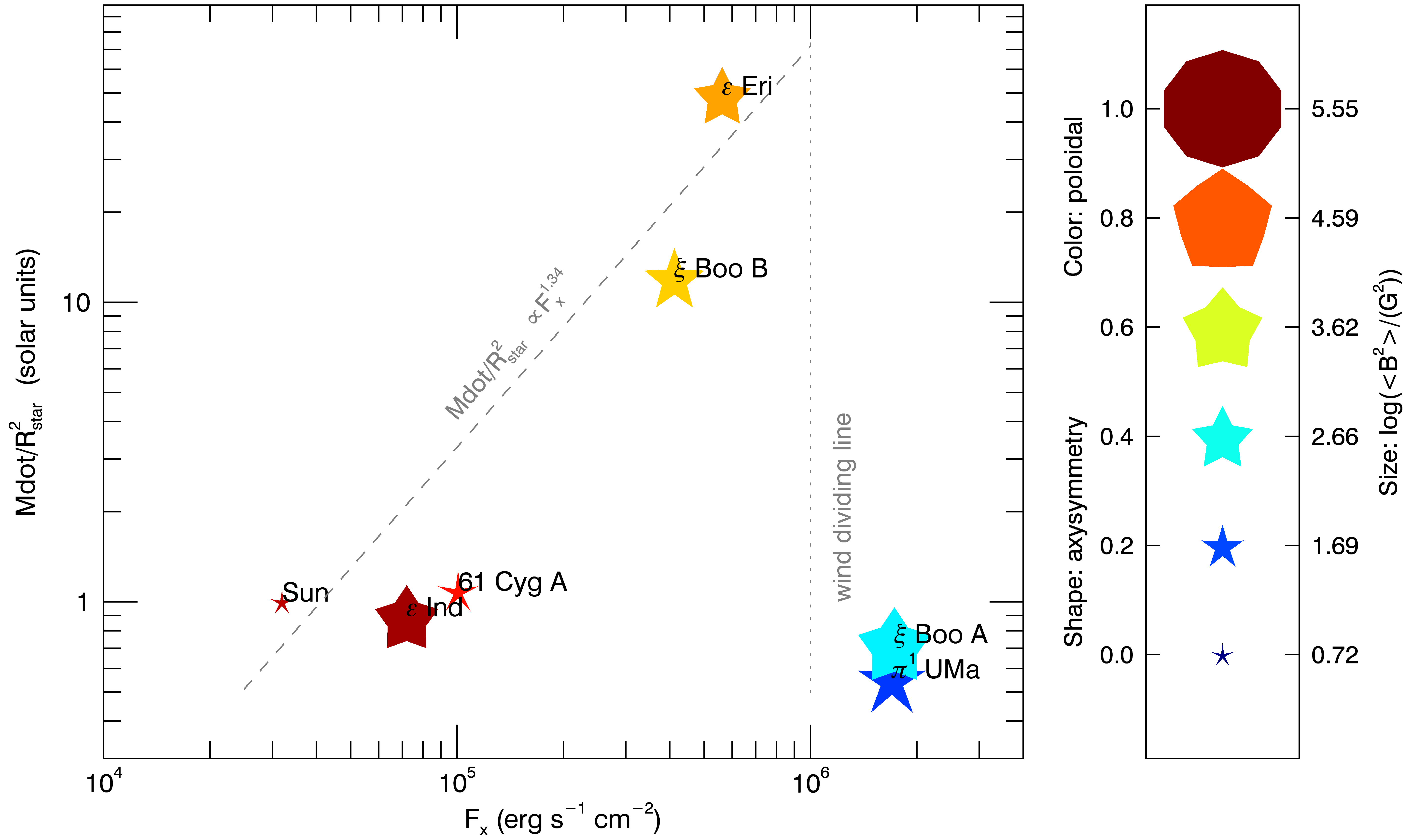}	
	\end{center}
	\caption{This diagram, kindly nick-named as {\it confusogram}, has been commonly used to summarise magnetic field topologies of a sample of stars \citep{2008MNRAS.390..545D,2008MNRAS.390..567M}. This is a five-dimension diagram. Here, the $y$ and $x$ axes represent the mass-loss rates per surface area versus X-ray fluxes. Each symbol represents one star in the sample of \citet{2014ApJ...781L..33W}  for which we have reconstructed surface fields (red points in Figure \ref{fig.wood14}). The size, shape and colour of the symbol represent the magnetic field characteristic of the star. Symbol sizes are proportional to $\log \langle B^2 \rangle$ (i.e., magnetic energy), their colours indicate the fractional poloidal energy (ranging from deep red for purely poloidal field $f_{\rm pol}=1$ to blue for purely toroidal field $f_{\rm pol}=0$), and their shapes indicate the fraction of axisymmetry of the poloidal component (ranging from a decagon for purely axisymmetric field $f_{\rm axi}=1$ to a point-shaped star for $f_{\rm axi}=0$). The $\mdot$ -- $F_x$ relation \citep{2004LRSP....1....2W}  is shown as a dashed line and the wind diving line at $F_x=10^6$~erg~cm$^{-2}$s$^{-1}$ is shown as a dotted line. The solar symbol was computed using the Sun-as-a-star magnetic map from \citet{2016MNRAS.459.1533V} for CR2109. Figure based on \citet{2016MNRAS.455L..52V}. 	\label{fig.confusogram}}
\end{figure*}

Observations of the solar wind with Ulysses spacecraft  revealed that the geometry of the solar magnetic field affects the velocity distribution of the solar wind \citep{2008GeoRL..3518103M}. When the Sun is in activity minimum and has a simpler magnetic field topology, close to a dipole, the wind structure is bi-modal, with large wind velocities in the (high-latitude) coronal holes than in the low-latitude region. On the other hand, when the Sun is in maximum activity, its magnetic field geometry becomes more complex, which is then reflected in the wind structure. Numerical simulations indeed show that the geometry of the stellar magnetic field affects the wind velocity \citep[e.g.][]{2009ApJ...699..441V,2014MNRAS.438.1162V}.

It is, therefore, natural to associate the break seen in the wind-activity relation \citep[][Figure \ref{fig.wood14}]{2004LRSP....1....2W} to changes in magnetic field topology. These authors showed that, for stars with X-ray fluxes $F_x \lesssim F_{x,6} = 10^{6}$ erg s$^{-1}$ cm$^{-2}$, the mass-loss rates were related to $F_x$ almost linearly. However, beyond $F_{x,6}$, this linear relation ceased to be valid. One of the suggestions was that, at $F_{x,6}$, the surface magnetic field topology suffers an abrupt change. For solar-like stars, $F_{x,6}$ roughly corresponds to an age of $600$~Myr, indicating that young stars could have mass-loss rates that are smaller than usually believed. 

Thanks to the several past and on-going ZDI surveys, some of which that I highlighted in Section \ref{sec.magnetism}, there are now several stars, from the sample observed by \citet{2014ApJ...781L..33W}, that have reconstructed magnetic maps. These stars are shown as red points in Figure \ref{fig.wood14} and they are: 61 Cyg A \citep{2016arXiv160601032B}, $\xi$ Boo A and B (\citealt{2012A&A...540A.138M}, Petit {\it et al.}, in prep.), $\epsilon$ Eri \citep{2014A&A...569A..79J}, $\pi^1$ UMa (Petit {\it et al.}, in prep.), $\epsilon$ Ind (Boisse {\it et al.}, in prep.) and EV Lac \citep{2008MNRAS.384...77M}. Their magnetic fields were analysed in light of the break in the wind-activity relation  \citep{2014ApJ...781L..33W} by \citet{2016MNRAS.455L..52V}. These authors showed that, although the stars to the right of the break (namely $\pi^1$ UMa and $\xi$ Boo A) have indeed more toroidal fields (Figure \ref{fig.confusogram}), the poloidal-toroidal transition in magnetic field topology from $F_x \lesssim F_{x,6}$ to  $F_x \gtrsim F_{x,6}$ seems to be very smooth. I.e., there is no sudden transition in magnetic topology at $F_{x,6}$, suggesting that the magnetic field topology alone can not explain the break in the mass-loss rate versus X-ray flux relation. Furthermore, it was also noted that the more active stars, which appear to have more toroidal fields (bluish symbols in Figure \ref{fig.confusogram}), present in general a variation of toroidal energy fractions with time. This means that the color of the symbols depicted in Figure \ref{fig.confusogram} for more active stars, such as $\pi^1$ UMa and $\xi$ Boo A, are likely to change in timescales of the order of years.

Since  $F_{x,6}$ roughly corresponds to an age of $600$~Myr, one might wonder how the topology of the sun evolved since its arrival in the main sequence. To trace the solar evolution, \citet{2016arXiv160503026R} studied the magnetic field topology of a sample of six solar twins, ranging in ages from 110 to 650 Myr.  Their results seem to confirm that there is indeed no sudden transition in magnetic field topology at $\sim$600 Myr. For their two oldest stars (BE Cet at 600 Myr and $\kappa^1$ Cet at 650 Myr), they noticed that the octupolar field was more prominent than in the younger solar twins. With two stars only showing the increase in octupolar field, it is hard to tell whether this is a real trend with age, especially because these young (active) stars are more likely to show magnetic field evolution in short timescales. The much older (few Gyr-old) solar twin 18 Sco, for example, was observed to be dominated by the quadrupolar field, with very small energies in the octupolar modes \citep{2008MNRAS.388...80P}.

\begin{figure}[ht]
	\centering
	\includegraphics[width=\columnwidth]{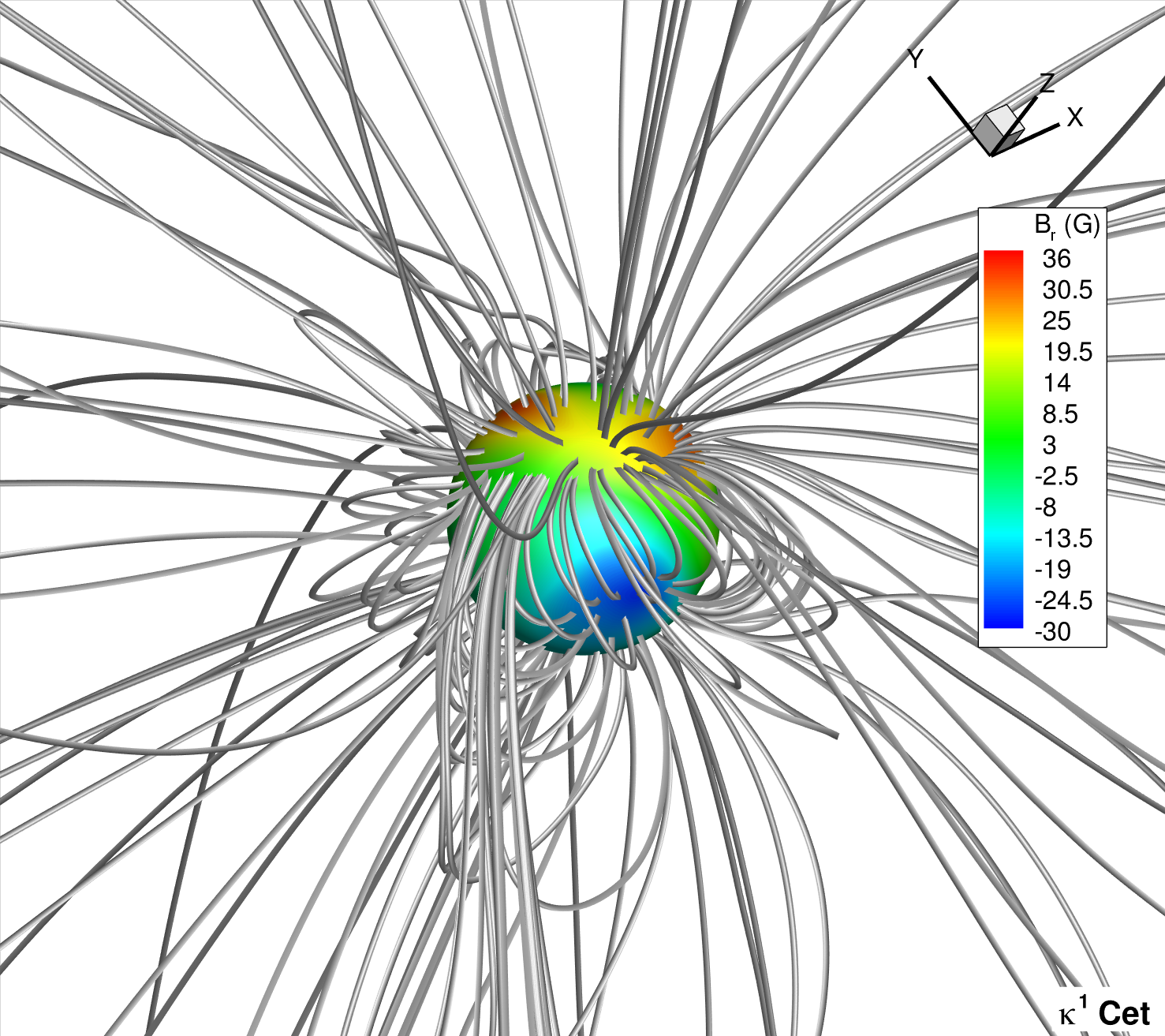}
	\caption{The magnetic field lines embedded in the wind of $\kappa^1$ Cet, a proxy for the young Sun when life arose on Earth. Figure adapted from \citet{2016ApJ...820L..15D}. \label{fig.3dlines}}
\end{figure}

Studying these solar twins can be quite illuminating if one wishes to recover the long-term history of the Sun and our solar system \citep{2003ApJ...594..561G,2005ApJ...622..680R,2007LRSP....4....3G,2014A&A...570A..99S,2016ApJ...817L..24A}. Recently, \citet{2016ApJ...820L..15D} presented a detailed study of the particle and magnetic environments surrounding $\kappa^1$ Cet, which has been recognised as a good proxy for the young Sun when life arose on Earth. From a set of spectropolarimetric observations, \citet{2016ApJ...820L..15D}  reconstructed the surface magnetic field of $\kappa^1$ Cet. This was then used in three-dimensional stellar wind simulations following the method of \citet[][Figure \ref{fig.3dlines}]{2015MNRAS.449.4117V}. They showed that the mass-loss rate of $\kappa^1$ Cet is about 50 times larger than the present-day Sun, in agreement also with other theoretical studies \citep{2016ApJ...817L..24A}. Due to this larger mass-loss rates, the ram pressure of the young solar wind impacting on the magnetosphere of the young Earth was larger than the present-day values. \citet{2016ApJ...820L..15D} showed that the magnetospheric sizes of the young Earth should have been reduced to approximately half the value it is today, or even smaller (a third), depending on the magnetic field intensity of the Earth:  similar to today's value or similar to the values estimated for the paleoarchean Earth \citep{2010Sci...327.1238T}, respectively. It is believed that smaller magnetospheric sizes could have an impact on atmospheric protection \citep{2013ApJ...770...23Z,2013A&A...557A..67V, 2014A&A...570A..99S}. 

Stellar wind studies have seen quite a stir in the recent years, with works having a broad range applications on different fields of Astrophysics: 
(a) three dimensional simulations of winds of planet-hosting stars \citep{2013MNRAS.436.2179L, 2012MNRAS.423.3285V,2015MNRAS.449.4117V, 2016A&A...588A..28A, 2016MNRAS.459.1907N}; (b) detailed studies of interactions between stars and close-in orbit planets \citep{2011ApJ...733...67C,2014ApJ...795...86S,2015ApJ...815..111S,2015A&A...578A...6M}; (c) effects of the magnetic field topology on the wind and angular momentum losses \citep{2013MNRAS.431..528J,2014MNRAS.438.1162V,2015ApJ...798..116R,2015ApJ...814...99R,2015ApJ...813...40G}; (d) evolution of the solar wind and its impact on the young Earth \citep{2014A&A...570A..99S,2016ApJ...817L..24A,2016ApJ...820L..15D}. I expect that this field will continue to grow in the next decade, as different research groups are expanding and ramifying. 
 
%%%%%%%%%%%%%%%%%%%%%%%%%%%%%%%%%%%%%%%%%%%%%%%%%%%%%%%%
\section{Effects of stellar magnetism and winds on planetary environments}\label{sec.planets}
The interplanetary medium is filled with stellar wind particles and magnetic fields. Compared to the Earth, the close-in location of hot-Jupiters orbiting solar-type stars implies that these exoplanets interact with 
\begin{enumerate}
\item  higher density external environment, 
\item higher ambient magnetic fields, and 
\item lower wind velocities, as close-in planets are likely located in the acceleration zone of the stellar wind. However, close-in planets also have higher orbital (Keplerian) velocities. The latter implies that, although stellar winds surrounding hot-Jupiter can present low velocities, the relative motion between the planet and the local wind can be quite large \citep[e.g.][]{2010ApJ...722L.168V}.
\end {enumerate}
 The environment surrounding close-in planets is, therefore, expected to be harsher than the environment surrounding any solar system planet.  

During epochs when the star is flaring or when coronal mass ejections are released more frequently, the environment surrounding these close-in planets can get even harsher. The radiative energy released in flares, for instance, can increase the energy input into the upper atmosphere of planets. As a consequence, this excess energy can heat the planetary atmosphere, which is then more likely to outflow, resulting in atmospheric escape. Coronal mass ejections, if directed towards an exoplanet, can cause an increase in the pressure of the quiescent wind surrounding the exoplanet. This, in turn, causes planetary magnetospheres to become smaller, which could enhance the polar cap area of the planet, through which atmospheric escape can happen. 

Recently, \citet{2012A&A...543L...4L} and \citet{2013A&A...551A..63B} reported temporal variations in the exosphere of HD\,189733b, a hot-Jupiter orbiting an early-K dwarf. Using transmission spectroscopy, these authors reported an excess absorption in the transit of  HD\,189733b in Ly-$\alpha$ line, as well as a longer transit duration than compared to another Ly-$\alpha$ transit at an earlier epoch (7 months before). Ly-$\alpha$ asymmetric transits, such as the one observed in HD\,189733b,  are usually explained as signatures of planet evaporation, in which the planet's atmosphere is leaking neutral hydrogen  \citep[e.g.,][]{2003Natur.422..143V,2014Sci...346..981K,2014MNRAS.438.1654V,2016A&A...591A.121B}. However, in the case of the Ly-$\alpha$ observations of HD\,189733b, the lack of an asymmetry  at a previous transit raised the following question: what is causing the increase/onset of atmospheric evaporation in HD\,189733b?
As it happens, 8 hours prior to the second transit observation, a flare from HD\,189733 was detected in X-ray observations. This led the authors of the study to propose that either the X-ray flare or its associated mass ejection (or both) was responsible for triggering (or increasing) the evaporation of HD\,189733b's atmosphere.

Since extreme ultraviolet and X-ray (UVX) emission of solar-type stars is observed to decrease with time, and UVX energy deposition on the top layers of planet atmospheres can cause evaporation, planets are more likely to undergo atmospheric evaporation at younger ages. Stellar age, nevertheless, is not the only ingredient that determines the UVX emission (i.e., activity state) of stars (this is why I say in Figure \ref{fig.wood14} that X-ray flux is a {\it rough} proxy for age). Rotation also matters. Activity indicators, such as X-ray (coronal) or CaII H\&K (chromospheric) emissions, are known to be linked to rotation. However, up to a certain age, roughly around 600 -- 700 Myr, stars at the same age can have a wide range of rotation periods and, therefore, a spread in activity indicators. It is only after stars become older that the age-rotation sequence tightens and one can directly infer ages from observed rotation rates \citep[e.g.][]{2003ApJ...586..464B,2013A&A...556A..36G,2015A&A...577A..28J}. The spread in rotation rates observed for young stars implies a spread in UVX luminosities, such that slow rotators at a given age will have low UVX luminosities than a fast-rotating star at the same age \citep{2015A&A...577L...3T}.

This implies that the history of the host star plays an essencial role in the evolution of atmospheric evaporation of their planets. \citet{2015A&A...577L...3T} showed that a 0.5-Earth-mass planet orbiting at 1 au around a solar-mass star will present an atmosphere with a very different hydrogen content at 4.5 Gyr (the age of the solar system), depending on whether the host star was a slow or fast rotator during its youth. Assuming an initial hydrogen atmosphere of $5\times 10^{-3}$ Earth mass, the atmosphere of the terrestrial planet is entirely lost before the system reaches 100 Myr-old if the host star begins its main-sequence evolution as a rapid rotator, while if the host star is a slow rotator, at 4.5 Gyr, the planet still has about 45\% of its initial atmosphere.

In the same way that stars affect the evolution of planets, planets can also affect the evolution of stars. In a series of works, \citet{2016A&A...591A..45P,2016arXiv160608027P} studied the effect planet engulfment would have in the evolution of 1.5 -- 2.5 $M_\odot$ stars. These authors focused on engulfment occurring at the red giant phase. After the stars evolve off of the main sequence, the radius expansion alters planet-star tidal forces and, as a consequence, giant planets (or sub-stellar objects) that were initially at large orbits (0.5 -- 1.5 au) can spiral inwards, and may eventually be engulfed by their host stars. During the migration and engulfment processes, the orbital angular momentum is transferred to the outer convective envelope of the red giant star, increasing its surface rotation velocities. \citet{2016arXiv160608027P} showed that the observed rotation velocities of several red giant branch stars cannot be explained by any reasonable model for single star evolution. They concluded that such stars were strong candidates to have  had recent planet-engulfment events. 

Similar to main-sequence stars \citep{2014MNRAS.441.2361V},  rotation is also intimately related to magnetism in evolved stars \citep{2015A&A...574A..90A}. One can then wonder whether an increase in rotation rate caused by engulfment events could also enhance stellar magnetism. This was the subject of the third study by \citet{privitera_c}, who concluded that the planet-induced magnetic field theory could be tested by observing the same fast rotating stars (i.e., the strong candidates of recent planet-engulfment events) to test whether they indeed possess measurable magnetic fields.

%%%%%%%%%%%%%%%%%%%%%%%%%%%%%%%%%%%%%%%%%%%%%%%%%%%%%%%%
\section{Concluding remarks}\label{sec.remarks}
Stellar winds play a central role in the evolution of cool stars. The intimate interplay between winds and magnetism enhance angular momentum loss, which have a profound impact in the evolution of stellar rotation, activity, magnetic field generation, internal structure, etc  (Figure \ref{fig:fig_big_picture}). As stars do not evolve in isolation, their surrounding planets are also affected by the changes their hosts undergo during their lifetime.

In this article, I reviewed some recent studies that focused on the mutual interplay between stellar magnetism, winds and exoplanets.  It is my personal belief that through the combined efforts between different communities, we can gather a more comprehensive view of the intricate evolution of stars and their planets.

%%%%%%%%%%%%%%%%%%%%%%%%%%%%%%%%%%%%%%%%%%%%%%%%%%%%%%%%
\section*{Acknowledgments}
{I would like to thank all the organisers of Cool Stars 19 for making it such a great conference and for the financial support that helped me attend this meeting. Finally, I thank the SAS pilots for giving me the opportunity to spend two extra days in Stockholm, with all my expenses paid for.}

\bibliographystyle{cs19proc}
\let\mnrasl=\mnras
\def\grl{{Geoph.~Research Letters}}

\end{document}